\documentclass[sigconf,nonacm]{acmart} 

\AtBeginDocument{%
  }

\setcopyright{acmlicensed}
\copyrightyear{2018}
\acmYear{2018}
\acmDOI{XXXXXXX.XXXXXXX}
\acmConference[Conference acronym 'XX]{Make sure to enter the correct
  conference title from your rights confirmation email}{June 03--05,
  2018}{Woodstock, NY}
\acmISBN{978-1-4503-XXXX-X/2018/06}

\acmSubmissionID{273} 

\usepackage{amsmath}
\usepackage{amsfonts}
\usepackage{algorithmic}
\usepackage{graphicx}
\usepackage{textcomp}
\usepackage{xcolor}

\usepackage[frozencache,cachedir=minted-cache]{minted}

\usepackage{multirow}
\usepackage{booktabs}
\usepackage{listings}
\usepackage{tabularx} 
\usepackage{CJKutf8}

\usepackage{csquotes}
\usepackage{float}
\usepackage[caption=false,font=footnotesize]{subfig}                          

\usepackage{footnote}
\makesavenoteenv{tabular}

\newcommand{\para}[1]{\medskip\noindent\textbf{#1}}

\newcommand{\eg}{\textit{e.g., }}
\newcommand{\ie}{\textit{i.e., }}

\newcommand{\pt}[1]{$#1\%$}

\newcommand{\figurevspace}{\vspace{-12pt}}
\newcommand{\tablevspace}{\vspace{-10pt}}

\begin{document}

\title{SPDZCoder: Teaching LLMs to Synthesize Privacy Computing Code without Massive Training Data}
\title{SPDZCoder: Combining Expert Knowledge with LLMs for Generating Privacy-Computing Code}

\author{Xiaoning Dong}
\email{dongxn20@mails.tsinghua.edu.cn}
\affiliation{%
  \institution{Tsinghua University}
  \city{Beijing}
  \country{China}
}

\author{Peilin Xin}
\email{xinpl21@mails.tsinghua.edu.cn}
\affiliation{%
  \institution{Tsinghua University}
  \city{Beijing}
  \country{China}}

\author{Jia Li}
\email{lijia@stu.pku.edu.cn}
\authornotemark[1]
\affiliation{%
  \institution{Peking University}
  \city{Beijing}
  \country{China}
}

\author{Wei Xu}
\email{weixu@tsinghua.edu.cn}
\authornote{Both authors equally contributed to advising this research.}
\affiliation{%
  \institution{Tsinghua University}
  \city{Beijing}
  \country{China}}

\renewcommand{\shortauthors}{Dong et al.} 


\begin{abstract}
Privacy computing receives increasing attention but writing privacy computing code remains challenging for developers due to limited library functions, necessitating function implementation from scratch, and data-oblivious requirement, contradicting intuitive thinking and usual practices of programmers. 
Automating the generation of privacy computing code with Large Language Models (LLMs) can streamline development effort and lower the barrier to using privacy computing frameworks since LLMs exhibit strong capabilities in coding tasks. However, existing LLMs still encounter challenges in code translation for privacy-preserving computation, such as translating Python to MP-SPDZ, due to the scarcity of MP-SPDZ data required for effective pre-training or fine-tuning.  Moreover, the lack of a standardized benchmark further complicates the evaluation of translation quality. 
To address the limitations, this work proposes SPDZCoder, a rule-based framework that combines LLMs with expert knowledge for generating privacy-computing code without requiring additional training data. 
Specifically, SPDZCoder employ a rigorous procedure for collecting high-quality expert knowledge to represent the semantic-expressing differences between Python and MP-SPDZ, and to derive transformation rules for translating Python to MP-SPDZ based on these knowledge. Then, SPDZCoder progressively converts Python code into MP-SPDZ code
using transformation rules in a three-stage pipeline.
To evaluate SPDZCoder, we manually constructed a benchmark dataset, SPDZEval, which comprises six data splits, each representing a distinct class of challenging tasks in MP-SPDZ implementation. Extensive experiments demonstrate that SPDZCoder achieves superior performance, significantly surpassing baselines in pass@1 and pass@2. Specifically, SPDZCoder attains an overall correctness of \pt{85.94} and \pt{92.01} in pass@1 and pass@2, respectively, whereas the best-performing baseline achieves only \pt{63.58} and \pt{76.36}, respectively. 



\end{abstract}


\maketitle

\section{Introduction} \label{sec:intro}
Multi-party computation (MPC) \cite{yao_protocols_1982, 10.1145/28395.28420} allows multiple parties to jointly compute a mutual function over their inputs and obtain the computation results without disclosing any participant’s private inputs. As a sub-field of cryptography, MPC receives increasing attention, and \textit{Multi-Protocol SPDZ} (\textit{MP-SPDZ}) \cite{keller_mp-spdz_2020} is a prevalent MPC framework that enables researchers to write programs achieving secure or privacy-preserving goals.

Despite the growing demand for privacy computing applications and frameworks (like MP-SPDZ), their adoption falls significantly short of expectations. 
One reason is the lack of skilled developers for privacy computing code as most developers rely on General Programming Languages (GPL) such as Python. Besides, the intricate features of MPC increase the cost to training them, \eg the data-oblivious requirement (discussed in Sec. \ref{subsec:bg-mpc}) contradicts usual practices of developers. 
Thus, this work aims to explore a question: \textit{can GPL code be automatically translated into MP-SPDZ code?}

Recently, Large Language Models (LLMs) \cite{touvron_llama_2023, touvron_llama_2023-1, claude3, openai_gpt-4_2023, gemini_team_gemini_2023, jiang_mixtral_2024, glm2024chatglm, gpt-4o, deepseekai2024deepseekv3technicalreport} have made remarkable advancements. 
Building on their success, code large language models (code LLMs) \cite{deepseekv2, hui2024qwen2, nijkamp_codegen_2023, guo_deepseek-coder_2024} are widely employed in code intelligence tasks, such as code translation \cite{yangExploringUnleashingPower2024, macedoInterTransLeveragingTransitive2024a}.
More recently, reasoning LLMs (RLLMs) \cite{deepseekai2025deepseekr1incentivizingreasoningcapability, qwq-32b-preview, openai-o1} further enhance LLMs’ logical inference capabilities, improving their effectiveness in code generation. 
These models have achieved leading performance across multiple benchmarks, owing to their strong comprehension of code semantics and advanced reasoning abilities. 

    \begin{figure}[tbp] 
        \centering
        \includegraphics[width=0.49\textwidth]{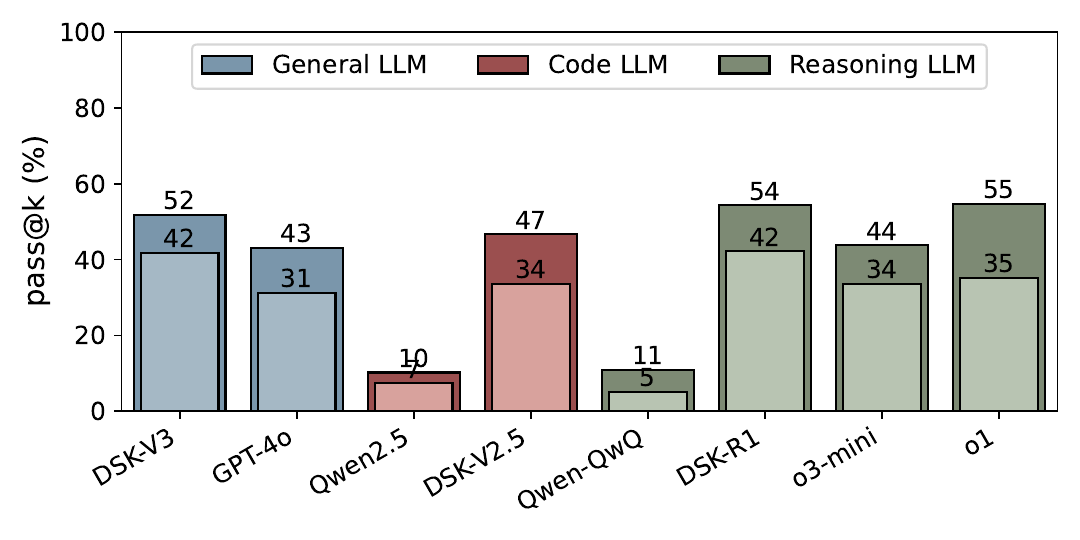} 
        \vspace{-27pt}
          \caption{Overall functional correctness (pass@1 and pass@2) of direct translation by the most recent advanced general, code and reasoning LLMs on SPDZEval (described in \ref{subsection:dataset}). Pass@1 is represented in light colors, while pass@2 is represented in dark colors. DSK stands for DeepSeek.}
        \Description[<short description>]{<long description>}
          \label{fig:direct-translation}
        \figurevspace
    \end{figure}

However, these LLMs struggle with generating privacy-preserving computation code, such as translating Python to MP-SPDZ. As shown in Figure \ref{fig:direct-translation}, OpenAI-o1 achieves only an overall pass@2 of \pt{55}. We attribute this to the scarcity of publicly available MP-SPDZ code, which is insufficient for LLMs to learn the significant differences between Python and MP-SPDZ even when they express the same semantics (\ie semantic-expressing differences).

As existing LLM-based methods, including the pre-training \cite{openai_gpt-4_2023,hui2024qwen2,deepseekai2025deepseekr1incentivizingreasoningcapability} and enhancing \cite{yangExploringUnleashingPower2024,macedoInterTransLeveragingTransitive2024a} paradigm have not explicitly addressed semantic-expressing differences when generating MP-SPDZ code from Python. To bridge this gap, we propose a rule-based approach that enables LLMs to address them by incorporating expert knowledge .
Our core idea is to collect high-quality expert knowledge to represent the semantic-expressing differences between Python and MP-SPDZ, and to derive transformation rules for translating Python to MP-SPDZ based on these knowledge. Thus, with the guidance of rules, we leverage the powerful LLMs to automatically translate Python code into MP-SPDZ code. We refer to this approach as SPDZCoder. SPDZCoder does not require additional training data and can be seamlessly applied to different LLMs.

There are two main challenges in implementing the SPDZCoder. The first challenge is \textit{how to develop rules for translation?} 
To address this challenge, we propose a rigorous procedure for collecting expert knowledge, consisting of three steps: (1) engaging MPC experts, (2) collaborating with them to identify semantic-expressing differences between Python and MP-SPDZ, and (3) deriving transformation rules based on these differences.  
Specifically, we categorize these differences into low-level and high-level ones. Low-level differences occur when Python data types, library functions, data structures, and methods have equivalent or approximate counterparts in MP-SPDZ but differ in naming or syntax (\ie misaligned names). In contrast, high-level differences include the data-oblivious requirement in MP-SPDZ and the absence of equivalent library functions compared to Python. 
To address high-level differences, we propose refactoring Python code into a simplified basic form, leading to a set of \textit{refactoring rules}. For low-level differences, we formulate a \textit{generation rule} that employs multiple in-context learning demonstrations to guide LLMs in replacing misaligned names and syntactic expressions accordingly.

The second challenge is \textit{how to combine expert knowledge with LLMs?} 
To address this, we propose a three-stage translation framework that progressively converts Python code into MP-SPDZ code using transformation rules. Specifically, SPDZCoder generates MP-SPDZ code through the following stages:

\textbf{Refactoring stage.} We apply a set of refactoring rules to refactor Python by creating the missing library functions with basic ones, refactoring non-oblivious statements such as $branch$ and $loop$ into oblivious form, and most importantly, replacing the non-oblivious algorithms with semantically-equivalent functions designed for privacy computing. 
We denote the refactored Python code to the \textit{Canonical Form Python code} (\textit{CFP}).

\textbf{Generation stage.} We leverage the generation rule to translation from CFP to MP-SPDZ code. As CFP only contains ``easy'' expressions, statements and functions to translate, we dynamically select applicable instructions and demonstrations in the generation rule, integrate them into a single prompt, and let LLM generate MP-SPDZ code from CFP in a single step. 
We also incorporate a self-reflection rule immediately after the generation process to refine the output and mitigate hallucinations.

\textbf{Repair Stage.} If test cases are available, we can execute the generated MP-SPDZ code with them and collect any execution error messages if the generated code is incorrect. We classify error messages into two categories: compilation/runtime errors (e.g., syntax errors and misused APIs) and logic errors (i.e., incorrect functionality). We then prompt the LLM with these error messages to fix any bugs or logical issues in the generated MP-SPDZ code, further improving performance.

To evaluate the effectiveness of SPDZCoder, we manually constructed a benchmark named SPDZEval, comprising 313 pairs of $\sf{\langle Python,\ MP\mbox{-}SPDZ \rangle}$ functions with test cases. Furthermore, SPDZEval is partitioned into six data splits, each representing a distinct class of challenging tasks in MP-SPDZ implementation. 
We compare SPDZCoder to the recently advanced LLM-based translation methods, including UniTrans \cite{yangExploringUnleashingPower2024} and InterTrans \cite{macedoInterTransLeveragingTransitive2024a}, as well as the state-of-the-art general, code and reasoning LLMs with MP-SPDZ documentation on SPDZEval. 
Extensive evaluation results show that SPDZCoder outperforms strong baselines in pass@1 and pass@2.  
Specifically, SPDZCoder achieves an overall functional correctness of \pt{85.94} and \pt{92.01} in pass@1 and pass@2, respectively, whereas the best-performing baseline achieves only \pt{63.58} and \pt{76.36}, respectively. 

To the best of our knowledge, SPDZCoder is the first to explore how to translate GPL (Python) to MP-SPDZ code. SPDZCoder explicitly teaches LLM the semantic-expressing differences at different levels by incorporating expert knowledge. The contribution of our work can be concluded as follows: 
\begin{itemize}
    \item We propose a rigorous procedure for collecting expert knowledge and formulate transformation rules from expert knowledge to address the semantic-expressing differences between Python and MP-SPDZ.
    \item We propose a novel rule-based framework, namely SPDZCoder that combines LLMs with expert knowledge to translate Python to MP-SPDZ.
    \item We construct a manually-written benchmark dataset containing over 300 pairs of $\sf{\langle Python,\ MP\mbox{-}SPDZ \rangle}$ functions with test cases for evaluating future approaches.
    \item We conducted extensive experiments to evaluate SPDZCoder against baselines. Evaluation results demonstrate the effectiveness of SPDZCoder in performing code translation for privacy computing. 
\end{itemize}
\section{Background}
In this section, we introduce the preliminary concepts necessary for understanding MPC.
\subsection{Multi-party Computation} \label{subsec:bg-mpc}
Multi-party computation enables multiple parties who do not trust each other or any common third party to securely compute a function over their private inputs. Concretely, all parties agree on a function to compute, and then running an MPC protocol to jointly compute the outputs of the function. All parties can only get the results of the computation while dishonest parties can not reveal the inputs of honest parties. A classical MPC (2PC) problem is Yao’s Millionaires Problem \cite{yao_protocols_1982} introduced by Andrew Yao in the 1980s.

MPC requires data-oblivious algorithms. Formally, an algorithm $\mathcal{A}$ is considered data-oblivious if, for any two inputs $x$ and $y$ of the same length, the sequence of instructions executed and the sequence of memory accesses performed by $\mathcal{A}$ on $x$ and $y$ are identical. 
When an algorithm is not data-oblivious, an adversary can potentially infer information about the private inputs by observing (tracking) the execution pattern. 
Data-oblivious algorithms ensure that the observable behavior of the computation (i.e., the sequence of executed instructions and the timing and locations of memory accesses) remains independent of the secret inputs, thereby preserving privacy. Not all algorithms have an oblivious counterpart, \eg there is no \emph{oblivious quick sort}.

\begin{table*}[tbp] 
  \centering
  \caption{High-level Semantic-Expressing Differences, Refactoring Rules and Their Descriptions.}
  \resizebox{\linewidth}{!}{
    \begin{tabular}{l|c|m{28.415em}} 
      \toprule
      \multicolumn{1}{l|}{\textbf{Differences}} & \textbf{Refactoring Rule Instances} & \multicolumn{1}{c}{\textbf{Description}} \\
      \midrule
      math and universal function & Linear\_NonLinear & Refactor advanced non-linear function into the combination of basic non-linear functions. \\
      \midrule
      \multirow{2}*{data structure and syntax sugar} & DataStructure & Eliminate advanced data structure method e.g. List.append(). \\
      \cmidrule{2-3}          & SyntaxSugar & Eliminate Python syntax sugars, e.g. ternary expression \\
      \midrule
      loop and iteration & RewriteWhileLoop & Refactor while loop into for loop if applicable. \\
      \midrule
      array operation & EliminateAdvancedArrayOperations & Break advanced Numpy array operations (array creation, indexing and manipulation) into simple operations.  \\
      \midrule
      \multirow{3}*{ocf\_break\_continue} & EliminateBreak & Eliminate break keyword and refactor into data-oblivious form \\
      \cmidrule{2-3}          & EliminateContinue & Eliminate continue keyword and refactor into data-oblivious form \\
      \midrule
      \multirow{4}*{ocf\_branch} & NestedIf\_MultipleReturn & Refactor into plain conditions (without nest condition statement) and single return statement \\
      \cmidrule{2-3}          & ChainedComparison & Refactor ChainedComparison into comparisons with logic operation \\
      \cmidrule{2-3}          & ObliviousForm & Change the code into data-oblivious form \\
      \bottomrule
    \end{tabular}%
    }
  \label{tab:high_rules}%
\end{table*}%

\subsection{MP-SPDZ framework}\label{subsec:mpspdz}

MP-SPDZ is a versatile programming framework for MPC, providing a high-level programming interface based on Python including data types, instructions, library functions and classes. 

MPC frameworks \cite{mohassel_aby3_2018,li_privpy_2019} differ significantly from Python in semantic implementation, and MP-SPDZ is no exception. 
First, the data types and supported arithmetic operations are not aligned. MP-SPDZ additionally provides secure and clear data types such as secrete int (\texttt{sint}), secrete fixed float (\texttt{sfix}), clear int ({\texttt{cint}) and clear fix (\texttt{cfix}), among others. On the one hand, MP-SPDZ uses fixed point data type and corresponding arithmetic operations instead of floating point data type due to efficiency concerns. On the other hand, each data type supports a limited set of arithmetic operations and has certain restrictions. For example, \texttt{sint} data does not support bitwise logic operation.

Second, memory access and control flow requires distinct implementation. 
For example, when the \textit{condition} in \mintinline{python}{if} statements is a secrete variable, the code should be written in a data-oblivious manner. Specifically, we represent the final result as the combination of results from each branch in order to achieve data-oblivious execution pattern.
In the following example, when the variable \mintinline{python}{x} is secret, the following code:
\begin{minted}[frame=single, fontsize=\footnotesize]{python}
# Assume x is a float number
from math import sqrt
if x>0: 
    y = sqrt(x)
else:   
    y = sqrt(-x)
\end{minted}
should be written in a data-oblivious way under MP-SPDZ platform as follows:
\begin{minted}[frame=single, fontsize=\footnotesize]{python}
# Assume x is a sfix number
from Compiler.mpc_math import sqrt
y = (x>0) * sqrt(x) + (x<=0) * sqrt(-x)
\end{minted}
We can find that no matter what the value of variable \mintinline{python}{x} is, the execution path remains unchanged. Note that one can also use built-in \mintinline{python}{y = (x>0).if_else(sqrt(x), sqrt(-x))} for such simple case.
Third, unlike the comprehensive algorithm libraries such as Numpy, MP-SPDZ merely provides a few basic APIs, requiring programmers to implement many advanced computation functions and manipulations on \texttt{Array, Matrix} from scratch. For example, MP-SPDZ only offers 7 basic trigonometric functions and 6 non-linear math functions in its \texttt{mpc\_math} module while Numpy provides 18 and 38 respectively.

\section{Approach}
    In this section, we first introduce the procedure for collecting expert knowledge (\ie the rule development process), followed by the details of the SPDZCoder pipeline.
    \subsection{Collecting Expert Knowledge}
    We propose a rigorous procedure for collecting expert knowledge to guide the translation process from Python to MP-SPDZ. We collaborate with MPC experts to identify the semantic-expressing differences between Python and MP-SPDZ, and formulate transformation rules for translation. The detail process consists of three key steps:
    
    \para{Engaging MPC Experts.}
    We engaged four MPC experts from one organization~\footnote{We disclose their names and affiliation after the double-blind review process.}, each with at least three years of experience in developing MPC frameworks and applications. Notably, one expert designed and developed the pre-commercial prototype of the organization's closed-source MPC framework. We engaged these experts through prior collaboration within their organization.
    
    \para{Summarizing Semantic-Expressing Differences.}
Through discussions with these experts, we identified the semantic-expressing differences between Python and MP-SPDZ (i.e., the challenges in translating Python to MP-SPDZ), particularly the unique constraints and conventions in expressing common semantics within MP-SPDZ.

Among these differences, some are merely misaligned names, meaning that data types, library functions, data structures, and methods in Python have equivalent or approximate counterparts in MP-SPDZ but differ in naming. For example, \mintinline{python}{numpy.power(x,y)} is semantically equivalent to \mintinline{python}{mpc_math.pow_fx(x,y)} in MP-SPDZ. We refer to these as low-level semantic-expressing differences.

The remaining differences stem from disparities in the richness of ways to express semantic or fundamental differences in computational paradigms, primarily involving the absence of equivalent library functions and the data-oblivious requirements of MPC.
For example, the \emph{condition} in \emph{branch} statements cannot include any secret data type to ensure data obliviousness, , and~\emph{quick sort} algorithm should be replaced with oblivious sorting algorithm (\eg oblivious radix sort\cite{hamada2014oblivious}) due to the non-existence of oblivious quick sort. We refer to these as high-level semantic-expressing differences.

    We conclude that high-level semantic differences encompass computational function library (e.g., nonlinear and trigonometric functions), advanced data structure method, syntax sugar, data obliviousness (\eg control flow, oblivious algorithm, memory access) and array / matrix operation (e.g., indexing, slicing, concatenation), and we present the summarization in Table~\ref{tab:high_rules}.
    
    Low-level differences include variations in the naming and syntax of data types, operators, control flow statements, computational functions, containers and their operations, and sorting algorithms compared to their plaintext counterparts

    \para{Deriving Rules Based on Semantic-Expressing Differences at Various Levels.}

    Recall that Python and MP-SPDZ exhibit distinct approaches for expressing semantics and the differences range from low to high levels. While deriving a rule to solve the low-level semantic issues is relatively simple, e.g., replacing the function name from \mintinline{python}{numpy.power(x,y)} to \mintinline{python}{mpc_math.pow_fx(x,y)}, the hard parts lies in how to define rules to address the data oblivious requirement and the absence of library functions. Thus, we separately derive rules to address the differences at their respective levels.

    To address high-level differences, we propose refactoring Python code into a simplified basic form, leading to a set of \textit{refactoring rules}. For example, to address the absence of many computational functions in MP-SPDZ, we leverage the fact that many nonlinear functions in MPC can be expressed using a small set of four fundamental nonlinear functions. 
    Based on this, we establish a rule to simplify the implementation of complex nonlinear functions. Similarly, high-level semantic expressing differences in operations (\ie method) associated with container data types are decomposed into fundamental ones, and complex Python syntactic constructs (e.g., syntactic sugar) are rewritten using basic statements. 
    Overall, we designed ten {refactoring rules} to cover and address the identified high-level semantic-expressing differences. 
    Table \ref{tab:high_rules} presents them and their descriptions related to certain high-level semantic-expressing discrepancy. 
    
    For low-level semantic-expressing differences, we initially created six rules, but instead of applying them progressively, we grouped them into one rule, named \textit{generation rule} to facilitate elinating all low-level semantic-expressing differences in a single step. 
    
    \subsection{Automatic Code Translation Pipeline}
    
    In this section, we describe how SPDZCoder leverages expert knowledge (\ie rules) to guide LLMs translate Python to MP-SPDZ. As depicted in Figure \ref{fig:method_pipeline}, the pipeline comprises three main stages: 
    (1) \textbf{Refactoring}: We apply a set of refactoring rules to prompt LLM to refactor Python code into CFP (discussed later).
    (2) \textbf{Generation}: We apply a generation rule to CFP to generate target MP-SPDZ code and incoporate a self-reflection component to minimize hallucination.
    (3) \textbf{Repair}: We utilize execution messages of test cases to further refine the translation. We provide a step-by-step example of translation by SPDZCoder and rule prompt templates in Supplementary Material Section \ref{app:translation-example} and \ref{app:prompt-template}. 
    
    \para{\textbf{Stage 1: Refactor to CFP from Source Code.}}
    Each refactoring rule specifies a target modification to a certain kind of high-level semantic-expressing discrepancy. 
    We prompt LLM with them one by one to perform Python-to-Python refactoring. 
    These rules, in this stage, implement the missing library functions with basic ones (\eg we implement all non-linear math function with four basic ones: \texttt{exp}, \texttt{ln} , \texttt{sqrt} and \texttt{invertsqrt}, which have equivalent functions in MP-SPDZ), convert non-oblivious statements, such as branch and loop, into oblivious form, and replace the non-oblivious algorithms with semantically-equivalent functions designed for privacy computing, \eg replacing \emph{quick sort} with \emph{radix sort}. After applying the rules, we obtain an intermediate Python code where high-level semantic discrepancies have been eliminated. The intermediate Python code is more rigid but explicit (\ie the used functions and data structures have equivalent part in MP-SPDZ), and data-oblivious, which reduces the difficulty of the subsequent generation of MP-SPDZ code. 
    We refer to it as \textit{Canonical Form Python code} (\textit{CFP}). 
    
    For example, one category of refactoring rules is the \texttt{ocf\_break\_ \\continue} rule (row 5 in Table \ref{tab:high_rules}), which contains two instances to guide the LLM to eliminate the \mintinline{python}{break} and \mintinline{python}{continue} keywords in a Python program, respectively, while preserving the original semantics. 
    The following code example contains a \texttt{for-loop} with a \mintinline{python}{break} keyword. 
\begin{minted}[frame=single, fontsize=\footnotesize]{python}
a = INIT_ARRAY # Assume a is a secrete Array
for i in range(len(a)):
    if a[i]>2:
        break
    a[i] += 1
\end{minted}
    
    To eliminate \mintinline{python}{break}, the rule inserts a \texttt{boolean} flag initialized with \texttt{False} before the code and uses the flag to simulate the \mintinline{python}{break} statement. The resulting CFP is as follows:
\begin{minted}[frame=single, breaklines, fontsize=\footnotesize]{python}
a = INIT_ARRAY
flag = False
for i in range(len(a)):
    flag = flag or (a[i]>2)
    a[i] = flag*a[i] + (1-flag)*(a[i]+1)
\end{minted}

    In the refactoring stage, another issue is that the input Python code does not necessarily include all summarized code patterns. 
    To address this, we propose pattern matching strategy to allow SPDZCoder dynamically select applicable refactoring rules. Specifically, we utilize AST-based static code analysis or leverage LLM-based code analysis to detect whether the input Python code contains a high-level semantic-expressing discrepancy which is described in refactoring rule instances. If pattern match succeeds, SPDZCoder will apply the rule to the Python code; otherwise, SPDZCoder skip the rule. 
    The patter match strategy prevents SPDZCoder from applying unnecessary rules, improving efficiency.

    \begin{figure*}[htb]
        \centering
        \includegraphics[width=0.98\textwidth]{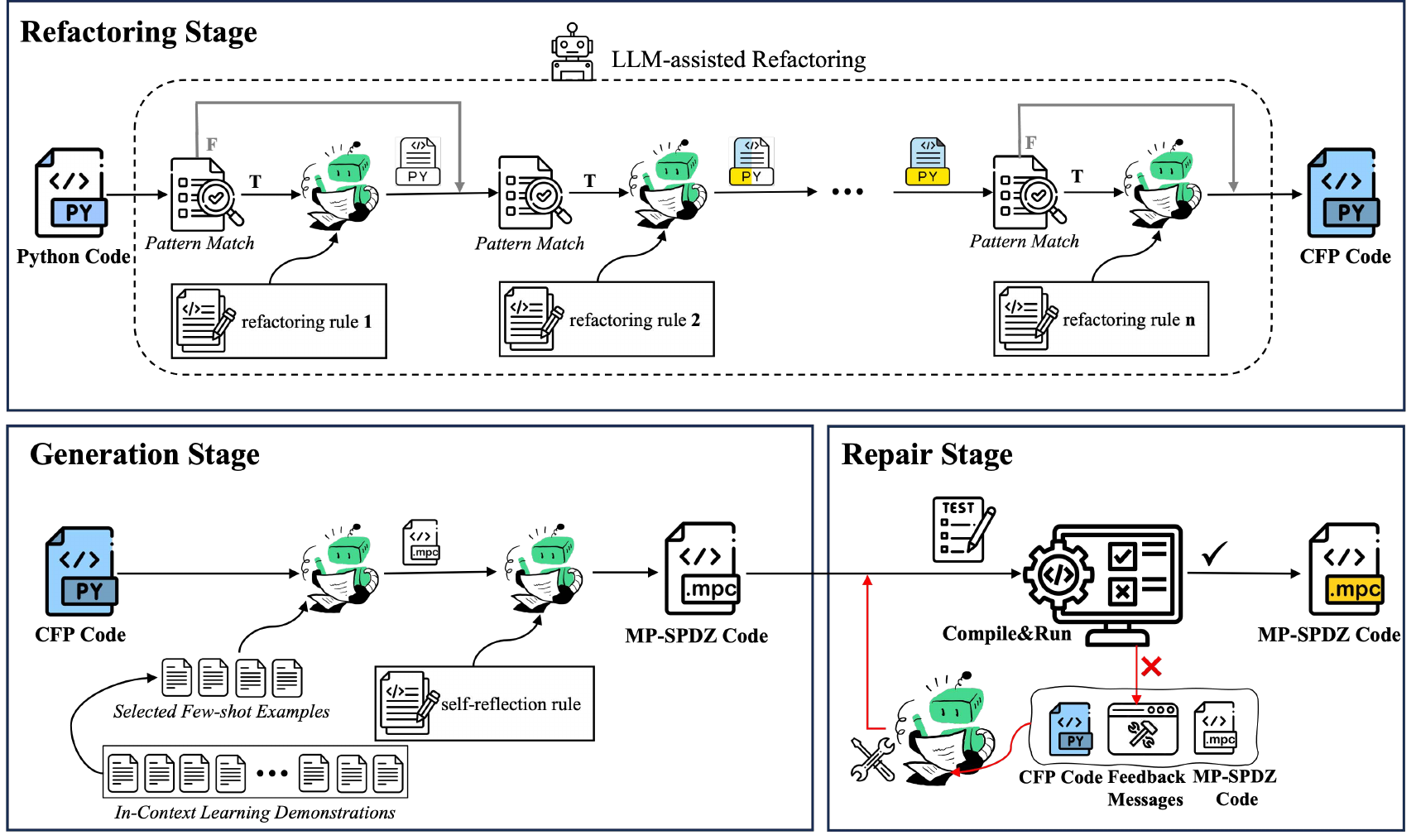} 
          \caption{Overview of SPDZCoder for automatic Python to MP-SPDZ code translation}
          \Description[<short description>]{<long description>}
          \label{fig:method_pipeline}
      \end{figure*}
    \para{\textbf{Stage 2: Translate to MP-SPDZ Code from CFP}}    
    In the generation stage, we apply a generation rule to instruct the LLM to translate CFP into MP-SPDZ. 
    Specifically, the generation rule consists of multiple in-context learning demonstrations addressing various low-level semantic-expressing differences, such as data types, logical operations, and function names. Each demonstration provides the knowledge of performing parallel conversion from CFP to MP-SPDZ. 
    For example, \texttt{float} data is translated in parallel to \texttt{sfix}; the logical operation \mintinline{python}{(x > -3 and x <= 0)} is translated in parallel to \mintinline{python}{(x > -3).bit_and(x <= 0)}; and the Numpy function \mintinline{python}{numpy.exp(x)} is translated in parallel to the mpc\_math function \mintinline{python}{mpc_math.pow_fx(cifx(math.e), x)}. 
    We employ pattern match to select applicable demonstrations and organize them into LLM chat messages so that LLM can regard them as chat history and refer them to perform translation. 
    
    We empirically find that buggy code can be generated due to hallucinations (an inherent limitation of LLMs), and it becomes more pronounced for generating MP-SPDZ code, presumably due to the absence of a common function library in MP-SPDZ and their insufficient knowledge of MP-SPDZ. 
    To address this, we incorporate a self-reflection \cite{ji-etal-2023-towards,10.1145/3712001} component immediately after the generation rule to alleviate hallucinations. 
    Specifically, the self-reflection rule prompts the LLM to check whether the operations, module names and function names are from the generation rule. If deviations are found, the LLM modifies the corresponding part of the code. 


    \para{\textbf{Stage 3: Compilation and Runtime Messages as Feedback}}

    In the repair stage, we optionally execute the generated MP-SPDZ code with test cases if they are available, and collect any execution error messages. The error messages indicate whether the code fails to execute due to compilation or runtime errors or contains logical errors (i.e., incorrect functionality). 
    
    If a compilation or runtime error occurs, we populate our \texttt{FixCom \\ -pilationRuntimeError} prompt template with $\sf{CFP}$, $\sf{error\ message}$, $\sf{generated\ MP\mbox{-}SPDZ\ code}$. If the code exhibits incorrect functionality, we fill our \texttt{FixFunctionalityError} prompt template with $\sf{CFP}$ and $\sf{generated\ MP\mbox{-}SPDZ\ code}$.
    We then employ these prompts and the selected few-shot demonstrations in the generation stage to prompt LLM to fix errors/defects by translating again. 
    The repair process repeats until the MP-SPDZ code is correct or the number of iterations reaches the predefined maximum retries ($\sf{max\_feedback}$). If the limit is reached, we accept the last rectified MP-SPDZ code as the final output.

\section{Experimental Setup}\label{sec:experiment}
    \subsection{Research Questions}
        \para{RQ1. What is the effectiveness of SPDZCoder compared to baselines?} 
        We evaluate SPDZCoder against three categories of LLM-based translation approaches on SPDZEval (discussed later). We also investigate the performance of SPDZCoder with various LLMs as backbone. 
        
        \para{RQ2. What is the effectiveness of refactoring and repair stage?} We study the impacts of the refactoring and repair stage in SPDZCoder on performance via ablation. 
        
        \para{RQ3. What is the correctness ratio of refactoring Python to CFP in SPDZCoder?} We investigate the refactoring correctness in refactoring stage since mis-refactoring can lead to performance drops in the generation stage.

        \para{RQ4. What are the token usage and savings achieved through rule pattern matching?} 
        We evaluate the average token consumption of SPDZCoder against baselines and, we examine the token savings by our pattern matching strategy in refactoring and generation stages, respectively.
    
    \subsection{SPDZCoder Setup} 
        \para{LLM Backbone} We employ a general LLM as the backbone for SPDZCoder, as our rule prompts contain both natural language (descriptions and instructions) and programming language (in-context learning demonstrations). Specifically, we adopt \texttt{gpt-4-turbo} (2024-04-09). 
        However, the LLM backbone can be replaced by any other LLMs at will \eg DeepSeek-V3, and we study the effectiveness of SPDZCoder using various LLMs in Section \ref{subsection:Eval-RQ1}.    

        \para{Generation Configuration and Hyper-parameters.}
        For the generation configuration of the backbone LLM, we set the ${\sf{temperature}}$ to $0.7$. A lower temperature yields more focused and deterministic outputs, making it more suitable for code generation, whereas the default value of $1.0$ is better suited for conversational tasks \cite{openaitemperature}. All other configurations remain unchanged.

        Our experiments involve two hyperparameters: $\sf{repitition}$ and $\sf{max\_feedback}$. The $\sf{repitition}$ parameter determines the number of MP-SPDZ code samples generated by our pipeline for each Python function in SPDZEval, while $\sf{max\_feedback}$ limits the maximum number of retries allowed in the repair stage. We set $\sf{repitition}$ to $2$ and $\sf{max\_feedback}$ to $3$.
        
    \subsection{Evaluation Dataset}
    \label{subsection:dataset}
        In this section, we introduce our created benchmark dataset SPDZEval.
        To ensure diversity in test functions, we selected an undergraduate student with a strong foundation in Python and experience in MP-SPDZ. Under the guidance of a development engineer, the student curated functions from Python practice websites \cite{w3resource,w3schools} and the NumPy library, covering a range of Python syntax, control flow and complexities, including basic arithmetic functions, array / matrix manipulations, and advanced mathematical functions.
        The undergraduate student was responsible for writing function docstrings, implementing the functions in Python and MP-SPDZ, designing test cases, and verifying correctness. The development engineer conducted a manual inspection of both the implementation and test cases to ensure reliability.
        Finally, we obtain $313$ pairs of $\sf{\langle source,\ target \rangle}$ (\ie $\sf{\langle Python,\ MP\mbox{-}SPDZ \rangle}$) functions in SPDZEval. The comparison of SPDZEval to commonly-used benchmarks are shown in Table \ref{tab:benchmark-stat}.
\begin{table}[htbp]
  \centering
  \Large
  \caption{Statistics of SPDZEval compared to representative code generation benchmarks. \#LOC represents lines of code.} 
  \resizebox{\linewidth}{!}{
    \begin{tabular}{lccccccc}
    \toprule
     \textbf{Dataset} &  \textbf{Language} &  \textbf{Construction} &  \textbf{Granularity} &  \textbf{Task} &  \textbf{\# Functions} &  \textbf{\#LOC} \\
    \midrule
     HumanEval \cite{chen_evaluating_2021} &  Python &  Manually &  Function-level &  Completion &  164   &  11.5 \\
     MBPP \cite{austin_program_2021} &  Python &  Manually &  Function-level &  Completion &  974   &  6.8 \\
     DS-1000 \cite{lai2022ds1000naturalreliablebenchmark} &   Python &  Automated &  Statement-level &  Completion &  1000  &  3.8 \\
    \midrule
    \midrule
     \textbf{SPDZEval} &  \textbf{MP-SPDZ} &  \textbf{Manually} &  \textbf{Function-level} &  \textbf{Translation} &  \textbf{313} &  \textbf{9.5} \\
    \bottomrule
    \end{tabular}%
    }
  \label{tab:benchmark-stat}%
\tablevspace
\end{table}%

        As shown in the table, SPDZEval is used to benchmark the code translation while others mainly focus on code completion.
        Moreover, SPDZEval is a function-level dataset, which is between statement-level granularity (rather simple) \cite{yinLearningMineAligned2018a,austin_program_2021,lai2022ds1000naturalreliablebenchmark} and class-level granularity (more complex) \cite{du_classeval_2023}. The function-level granularity means that we ask LLMs to generate one single code unit (\ie function) in a standalone way.
\begin{table}[htpb]
  \centering
\tablevspace
  \caption{Overview of each data split in SDPZEval}
  \resizebox{\linewidth}{!}{\Large
    \begin{tabular}{c|c|c|m{22.085em}} 
    \toprule
    \textbf{Split Name} & \textbf{Abbr.} & \textbf{\# Entry} & \textbf{Description of Code Snippet (Challenging Tasks)} \\
    \midrule
    array  & AR & 43    & Array access or traverse, including list indexing, slicing, iteration and access. \\
    \midrule
    loop & LO & 97    & Loop, iteration, multiple if, break and continue, multiple return. \\
    \midrule
    branch & BR & 34    & Multiple kinds of branch, e.g. nested if, multiple if, multiple return.  \\
    \midrule
    math & MA & 38    & Python scalar math functions and computations \\
    \midrule
    numpy & NP & 66    & Numpy array universal functions and Numpy array operation, \eg creation, indexing, manipulation. \\
    \midrule
    syntax & SX & 35    & Advanced syntax or syntax sugars in Python and built-in sophisticated functions, e.g. map(). \\
    \bottomrule
    \end{tabular}%
    }
    \label{tab:benchmark}%
\end{table}%

         We divide SPDZEval into six data splits and introduce them in Table \ref{tab:benchmark}. As shown in the table, each data split contains specific code patterns and can be used to assess the approach's translation capability to deal with the code pattern in the description (column four). For example, the numpy data split (row five) is related to array creation, manipulation and numpy universal functions (\ie ndarray math functions), and it can be used to evaluate the performance among different methods when the source code involves these operations. 

\begin{table*}[tbp]
  \centering
  \caption{Functional correctness (pass@1 and pass@2) of SPDZCoder vs. baselines across six data splits on SPDZEval. Results are presented in percentage format. The best results are highlighted in \textbf{bold}.} 
  \resizebox{\linewidth}{!}{
    \begin{tabular}{cc|c|c|c|c|c|c|c||c|c|c|c|c|c|c}
    \toprule
    \multicolumn{2}{c|}{\multirow{2}[4]{*}{method}} & \multicolumn{7}{c||}{pass@1}                          & \multicolumn{7}{c}{pass@2} \\
\cmidrule{5-7}\cmidrule{12-14}    \multicolumn{2}{c|}{} & \multicolumn{1}{c}{array} & \multicolumn{1}{c}{loop} & \multicolumn{1}{c}{branch} & \multicolumn{1}{c}{math} & \multicolumn{1}{c}{numpy} & \multicolumn{1}{c}{syntax} & \textbf{overall} & \multicolumn{1}{c}{array} & \multicolumn{1}{c}{loop} & \multicolumn{1}{c}{branch} & \multicolumn{1}{c}{math} & \multicolumn{1}{c}{numpy} & \multicolumn{1}{c}{syntax} & \textbf{overall} \\
    \midrule
    \multirow{12}[1]{*}{API docs} & \multicolumn{1}{l|}{GPT-4} & 44.19 & 24.74 & 26.47 & 10.53 & 39.39 & 37.14 & 30.35 & 60.47 & 53.61 & 32.35 & 26.32 & 53.03 & 62.86 & 49.84 \\
          & \multicolumn{1}{l|}{DeepSeek-V3} & 51.16 & 32.99 & 35.29 & 10.53 & 34.85 & 34.29 & 33.55 & 60.47 & 53.61 & 55.88 & 26.32 & 40.91 & 42.86 & 47.61 \\
          & \multicolumn{1}{l|}{GLM-4} & 34.88 & 21.65 & 5.88  & 23.68 & 21.21 & 17.14 & 21.4  & 55.81 & 22.68 & 20.59 & 31.58 & 34.85 & 25.71 & 30.99 \\
          & \multicolumn{1}{l|}{Qwen2.5-Coder-32B} & 72.09 & 47.42 & 44.12 & 28.95 & 34.85 & 45.71 & 45.37 & 83.72 & 63.92 & 64.71 & 36.84 & 43.94 & 62.86 & 59.11 \\
          & \multicolumn{1}{l|}{Qwen2.5-Coder-7B} & 48.84 & 15.46 & 17.65 & 18.42 & 15.15 & 22.86 & 21.41 & 62.79 & 27.84 & 26.47 & 39.47 & 19.7  & 31.43 & 32.59 \\
          & \multicolumn{1}{l|}{DeepSeek-V2.5} & 79.07 & 52.58 & 55.88 & 47.37 & 45.46 & 62.86 & 55.59 & 1.4   & 60.82 & 64.71 & 50    & 51.51 & 65.71 & 61.34 \\
          & \multicolumn{1}{l|}{DeepSeekCoder-33B} & 37.21 & 21.65 & 20.59 & 7.89  & 16.67 & 11.43 & 19.81 & 48.84 & 34.02 & 26.47 & 15.79 & 25.76 & 31.43 & 30.99 \\
          & \multicolumn{1}{l|}{OpenAI-o1} & 69.77 & 60.82 & 79.41 & 65.79 & 53.03 & 65.71 & 63.58 & 81.4  & 80.41 & 97.06 & 65.79 & 62.12 & 77.14 & 76.36 \\
          & \multicolumn{1}{l|}{OpenAI-o3-mini} & 58.14 & 71.13 & 79.41 & 55.26 & 57.57 & 60    & 64.21 & 67.44 & 81.44 & 85.29 & 73.68 & 66.67 & 68.57 & 74.44 \\
          & \multicolumn{1}{l|}{DeepSeek-R1} & 55.81 & 46.39 & 88.24 & 60.53 & 45.45 & 54.29 & 54.63 & 79.07 & 71.13 & 94.12 & 73.68 & 60.6  & 77.14 & 73.48 \\
          & \multicolumn{1}{l|}{Qwen-QwQ} & 34.88 & 22.68 & 50    & 31.58 & 28.79 & 17.14 & 29.07 & 48.84 & 35.05 & 79.41 & 44.74 & 39.39 & 31.43 & 43.45 \\
    \midrule
    \multirow{3}[1]{*}{UniTrans} & \multicolumn{1}{l|}{DeepSeek-V3} & 51.16 & 40.21 & 26.47 & 23.68 & 42.42 & 48.57 & 39.62 & 69.77 & 52.58 & 47.06 & 31.58 & 51.51 & 57.14 & 52.08 \\
          & \multicolumn{1}{l|}{DeepSeek-V2.5} & 48.84 & 34.02 & 26.47 & 10.53 & 31.82 & 34.29 & 31.95 & 62.79 & 44.33 & 50    & 23.68 & 40.91 & 51.43 & 45.05 \\
          & \multicolumn{1}{l|}{OpenAI-o1} & 34.88 & 47.42 & 44.12 & 23.68 & 33.33 & 28.57 & 37.38 & 51.16 & 64.95 & 61.76 & 34.21 & 46.97 & 45.71 & 53.03 \\
    \midrule
    \multirow{3}[2]{*}{InterTrans} & \multicolumn{1}{l|}{DeepSeek-V3} & 55.81 & 43.3  & 35.29 & 26.32 & 39.39 & 45.71 & 41.53 & 67.44 & 53.61 & 52.94 & 36.84 & 51.52 & 54.29 & 53.04 \\
          & \multicolumn{1}{l|}{DeepSeek-V2.5} & 46.51 & 32.99 & 29.41 & 13.16 & 28.78 & 28.57 & 30.67 & 60.47 & 43.3  & 55.88 & 23.68 & 43.93 & 42.86 & 44.73 \\
          & \multicolumn{1}{l|}{OpenAI-o1} & 46.51 & 48.45 & 50    & 31.58 & 27.27 & 42.86 & 41.2  & 60.47 & 63.92 & 61.76 & 42.11 & 42.43 & 57.14 & 55.27 \\
    \midrule
    \multicolumn{1}{c}{SPDZCoder } & \multicolumn{1}{l|}{GPT-4}  & \textbf{83.72} & \textbf{84.54} & \textbf{100} & \textbf{94.74} & \textbf{83.33} & \textbf{74.29} & \textbf{85.94} & \textbf{93.02} & \textbf{92.78} & \textbf{100} & \textbf{97.37} & \textbf{86.36} & \textbf{85.71} & \textbf{92.01} \\
    \bottomrule
    \end{tabular}%
    }
  \label{tab:main_result}%
\tablevspace
\end{table*}%

    \subsection{Baselines and Metric} \label{subsec:baselines_metric}
        \subsubsection{Baselines.} 
        As depicted in Figure \ref{fig:direct-translation} (see Introduction), existing LLMs fail to directly translate Python to MP-SPDZ program. To evaluate SPDZCoder, we design or adopt three approaches as strong baselines.
        
        \para{API Documentation (API Doc).}
        We enhance translation by prompting LLMs with MP-SPDZ API documentation and explicitly reminding them to be aware of semantic-expressing differences in the prompt. 
        For general LLMs, we additionally instruct them to summarize the source programming code (Python) before translation, placing the code summary after the MP-SPDZ API documentation in the prompt. 
        However, we do not apply this to code LLMs and reasoning LLMs, as code LLMs are less adept at generating natural language, and reasoning LLMs inherently follow a think-before-answer paradigm.
        The detailed prompt template is provided in Supplementary Material Section ~\ref{app:prompt-baseline}
        
        \para{UniTrans.}
        UniTrans \cite{yangExploringUnleashingPower2024} enhances LLM-based translation by augmenting the translation with LLM-generated test cases. It employs an error analyzer to extract error information—including error lines and error messages—from the execution results of incorrectly translated programs. The error information serves as hints to guide LLMs in rectifying incorrect programs.
        
        \para{InterTrans.}
        Intertrans \cite{macedoInterTransLeveragingTransitive2024a} leverages the multilingual capabilities of LLMs to enhance code translation via transitive intermediate translations.  
        Intertrans first employs a collection of intermediate programming languages and utilizes a planning algorithm (ToCT) to generate candidate translation paths (source-target or source-intermediate-target). 
        It then sequentially executes translation paths and validate the correctness of translated program through test cases, enabling early termination. Consistent with their setting, we employ Rust, JS, C++ and Go as intermediate programming languages.
        
        The LLM backbones for baselines are chosen as follows: For API Doc, we employ the most recently and powerful general, code and reasoning LLMs. Specifically we adopt GPT-4, DeepSeek-V3, GLM-4 (general LLMs); Qwen2.5-Coder-7,32B, DeepSeek-V2.5, DeepSeek Coder-33B (code LLMs); and OpenAI-o1,o3-mini, DeepSeek-R1, Qwen-QwQ (reasoning LLMs). 
        For UniTrans \cite{yangExploringUnleashingPower2024} and InterTrans \cite{macedoInterTransLeveragingTransitive2024a}, we replace their LLM backbones to the advanced ones: DeepSeek-V3, DeepSeek-V2.5, OpenAI-o1.
        
        
        \subsubsection{Evaluation Metric.}
        The pass@k metric \cite{NEURIPS2019_7298332f} measures functional correctness, where $k$ code samples are generated per problem, and if any sample in $k$ samples passes a set of given unit tests, the problem is deemed solved \cite{chen_evaluating_2021}. The metric reports the proportion of problems successfully solved. Consistent with mainstream works \cite{chen_evaluating_2021,austin_program_2021,liu_is_2023,du_classeval_2023,yu_codereval_2024}, we adopt it to evaluate whether the generated MP-SPDZ codes are functionally correct. Specifically, we evaluate the \textbf{pass@1}  and \textbf{pass@2} metrics of SPDZCoder on SPDZEval, respectively.

\section{Evaluation and Analysis} 
    In this section, we comprehensively evaluate SPDZCoder and report the experimental results of the research questions.
    \subsection{\textbf{RQ1. Effectiveness of SPDZCoder.}}\label{subsection:Eval-RQ1}
    We first evaluate the effectiveness of SPDZCoder against strong baselines that employ general, code and reasoning LLMs as backbones on SPDZEval, with results shown in Table \ref{tab:main_result}. Our observations are three fold: 
    
    \textbf{(1) SPDZCoder surpasses the most recently advanced baselines in pass@1 and pass@2 by a large margin.} Specifically, SPDZCoder achieves an overall correctness of \pt{85.94} and \pt{92.01} for pass@1 and pass@2, respectively, whereas the best-performing baseline attains \pt{63.58} and \pt{76.36}.
    
    The superior effectiveness of SPDZCoder over baselines can be attributed to the following two aspects. 
    First, SPDZCoder decomposes the translation task into a refactoring stage and a generation stage, where semantic-expressing differences at different levels are separately addressed.
    In particular, the refactoring stage reduces the difficulty of the following generation process, as the CFP code consists solely of simple and basic statements and functions while meeting the data-oblivious requirement. This allows the target MP-SPDZ code to be directly generated in a single attempt. 
    Second, SPDZCoder incorporates execution messages from test cases as feedback, guiding LLMs to rectify incorrectly translated programs (discussed in Section \ref{subsection:Eval-RQ2}).

    \textbf{(2) Accompany the results of direct translation shown in Figure \ref{fig:direct-translation} (see Introduction), we find that API Doc provides a modest enhancement for LLMs.} 
    For example, API Doc, with reasoning LLM, Deepseek-R1, as its backbone, achieves an overall pass@1 and pass@2 of 64\% and 75\%, respectively.  
    This improvement can be attributed to the fact that API documentation introduces MP-SPDZ knowledge to LLMs in a simple, coarse-grained manner via in-context learning, may indicating that effectively providing MP-SPDZ knowledge to LLMs is important for translation. 

    \textbf{(3) Similarly, accompany with Figure \ref{fig:direct-translation}, we observe that UniTrans and InterTrans almost fail to enhance LLMs.}
    For example, InterTrans, with OpenAI-o1 as the backbone, yields only an absoulte improvement of 6.05\% and 0.63\% in pass@1 and pass@2, respectively. 
    We conjecture that the reason is these approaches primarily involve widely used programming languages in plaintext computing, which have extensive online resources available for training.
    As a result, LLMs have been well-trained on these languages and already exhibit strong baseline performance. Building on this foundation, these methods can further enhance translation performance. 
    However, the availability of MP-SPDZ data is limited. Moreover, translating from Python to MP-SPDZ not only involves the inherent discrepancy between the two programming languages but also introduces semantic-expressing differences between plaintext and secure computation. These challenges prevent existing LLMs from effectively performing code translation in privacy computing scenarios.
    
    \medskip
    We further evaluate SPDZCoder's efficacy when using various LLMs as backbone. As shown in Table \ref{tab:various-backbone}, SPDZCoder consistently maintains its performance. Specifically, using DeepSeek-R1 as its backbone, SPDZCoder attains an overall correctness of \pt{88.18} and \pt{92.97} in pass@1 and pass@2, respectively. Replacing the backbone with a considerably weaker model, \eg GLM-4, results in an absolute drop of \pt{11.5} and \pt{10.54} in pass@1 and pass@2, while still outperforms baselines with state-of-the-art LLMs as backbone.
\begin{table}[htbp]
  \centering
  \caption{Overall pass@1 and pass@2 of SPDZCoder using various LLMs as backbone.}
  \setlength{\tabcolsep}{15pt}  
  {\renewcommand{\arraystretch}{1.0} 
  \scalebox{0.9}{
    \begin{tabular}{lcc}
    \toprule
    \multicolumn{1}{c}{\multirow{2}[4]{*}{SPDZCoder}} & pass@1 & pass@2 \\
\cmidrule{2-3}          & 85.94 & 92.01 \\
    \midrule
    \textit{with} GLM-4 & 74.44 & 81.47 \\
    \textit{with} DeepSeek-V3 & 85.63 & 91.05 \\
    \textit{with} DeepSeek-R1 & 88.18 & 92.97 \\
    \bottomrule
    \end{tabular}%
    }
}
  \label{tab:various-backbone}%
\end{table}%

    Finally, SPDZCoder exhibits room for improvement on the numpy and syntax data splits in SPDZEval, achieving an overall pass@2 of \pt{86.36} and \pt{85.71}, respectively.
    Figure \ref{fig:rq1_examples-1} presents an example of QR decomposition. As shown, SPDZCoder either references a non-existent MP-SPDZ \texttt{Matrix} method or directly inserts a placeholder function. 
    One possible reason is that our rules may not fully cover the high-level semantic differences in Numpy or advanced Python syntax, due to the rich features of advanced Numpy ndarray operations, such as integer array indexing. Another potential factor is that these tasks may be inherently too complex for SPDZCoder.
    \begin{figure}[htbp]
    \centering
    \includegraphics[width=\linewidth]{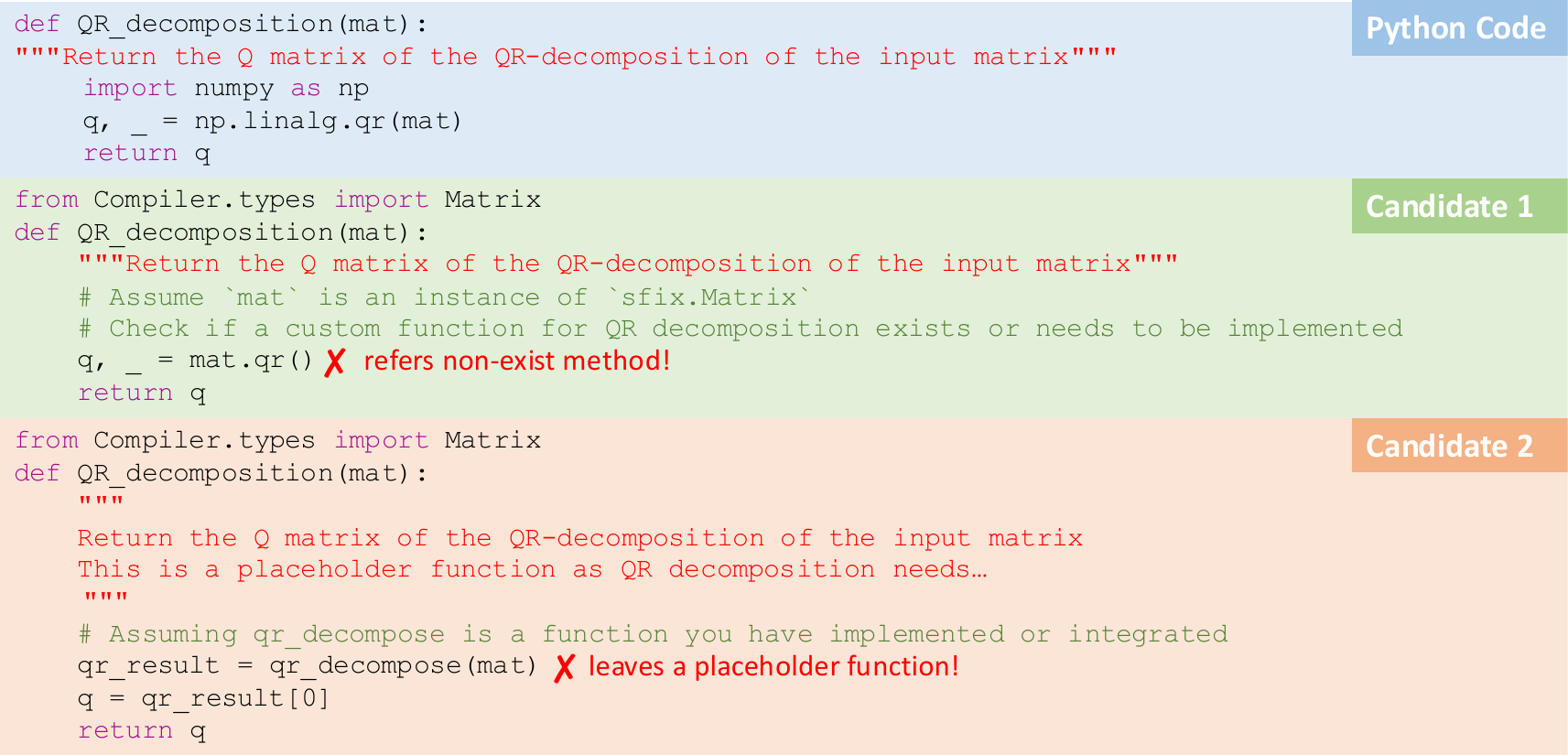} 
      \caption{QR decomposition: a difficult example for SPDZCoder.}
      \Description[<short description>]{<long description>}
      \label{fig:rq1_examples-1}
      \figurevspace
\end{figure}


    \subsection{\textbf{RQ2. Effectiveness of the Refactoring and Repair Components.}}
    \label{subsection:Eval-RQ2} 
\begin{table*}[htbp]
  \centering
  \caption{Correctness of SPDZCoder without refactoring and repair. Performance drop is calculated as the absolute difference.}
  \resizebox{\linewidth}{!}{
    \begin{tabular}{l|c|c|c|c|c|c|c||c|c|c|c|c|c|c}
    \toprule
    \multicolumn{1}{c|}{\multirow{2}[4]{*}{Setting}} & \multicolumn{7}{c||}{pass@1}                          & \multicolumn{7}{c}{pass@2} \\
\cmidrule{5-6}\cmidrule{12-13}          & \multicolumn{1}{c}{array} & \multicolumn{1}{c}{loop} & \multicolumn{1}{c}{branch} & \multicolumn{1}{c}{math} & \multicolumn{1}{c}{numpy} & \multicolumn{1}{c}{syntax} & overall & \multicolumn{1}{c}{array} & \multicolumn{1}{c}{loop} & \multicolumn{1}{c}{branch} & \multicolumn{1}{c}{math} & \multicolumn{1}{c}{numpy} & \multicolumn{1}{c}{syntax} & overall \\
    \midrule
    SPDZCoder & \textbf{83.72} & \textbf{84.54} & \textbf{100} & \textbf{94.74} & \textbf{83.33} & \textbf{74.29} & \textbf{85.94} & \textbf{93.02} & \textbf{92.78} & \textbf{100} & \textbf{97.37} & \textbf{86.36} & \textbf{85.71} & \textbf{92.01} \\
    \textit{w/o} refactoring & 58.14 & 29.9  & 20.59 & 21.05 & 34.85 & 34.29 & 33.23 (\textcolor{red}{\pt{52.7}$\downarrow$}) & 65.12 & 41.24 & 41.18 & 31.58 & 43.94 & 54.29 & 45.37 (\textcolor{red}{\pt{46.6}$\downarrow$}) \\
    \textit{w/o} repair & 74.42 & 72.16 & 85.29 & 78.95 & 71.21 & 60    & 73.16 (\textcolor{red}{\pt{12.8}$\downarrow$}) & 90.7  & 84.54 & 85.29 & 92.11 & 78.79 & 68.57 & 83.39 (\textcolor{red}{\pt{8.6}$\downarrow$}) \\
    \bottomrule
    \end{tabular}%
    }
  \label{tab:ablation}%
\end{table*}%

    We first examine the impact of SPDZCoder's refactoring stages on translation through ablation, with results shown in Table \ref{tab:ablation}. 
    From the table, we observe that the overall correctness drops drastically to 33.23\% and 45.37\% in pass@1 and pass@2, respectively, highlighting the importance of separately addressing semantic-expressing differences at various levels in SPDZCoder. 

    Next, we investigate the effectiveness of SPDZCoder's repair stage. 
    As shown in Table \ref{tab:ablation}, the repair stage improves the overall correctness from \pt{73.16} and \pt{83.39} to \pt{85.94} and \pt{92.01}, yielding absolute improvements of \pt{12.78} and \pt{8.62} in pass@1 and pass@2, respectively. The results indicate that auxiliary information from run-time message contains feedback information for SPDZCoder to rectify its synthesized programs. 
    Additionally, we observe that SPDZCoder maintains its superior performance over baselines without the repair stage. 

    \begin{figure}[htbp]
    \centering
    \includegraphics[width=\linewidth]{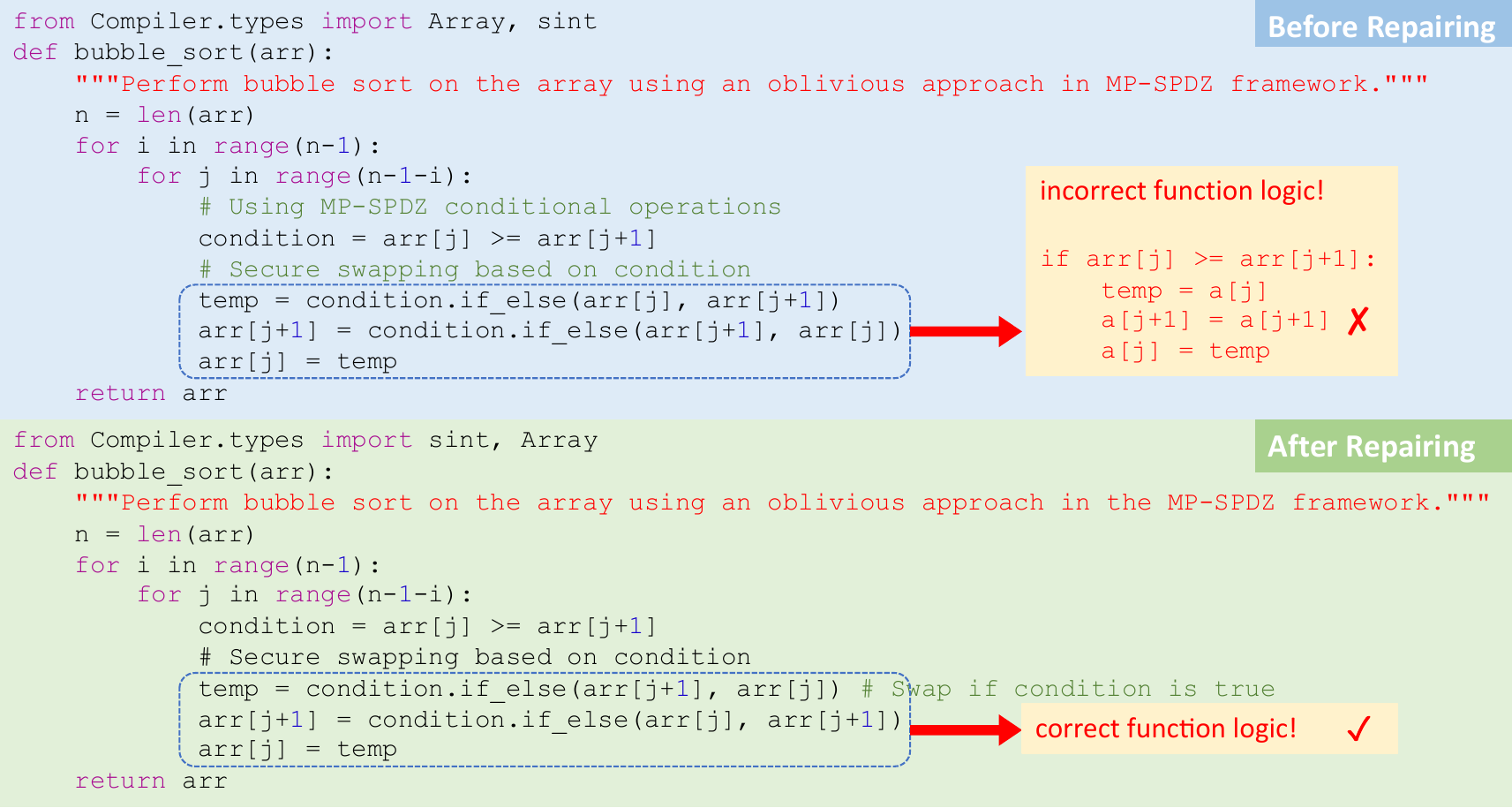} 
      \caption{Example of the repair component correcting the faulty logic in a bubble sort function.}
      \Description[<short description>]{<long description>}
      \label{fig:rq2_feedback_examples-1}
\end{figure}

    Figure \ref{fig:rq2_feedback_examples-1} illustrates how the feedback component rectify incorrect function logic. In this example, although no compilation errors occur, the feedback component identifies the faulty logic in its generated \texttt{bubble\_sort} function, and successfully corrects its functionality by adjusting the assignment statement. 
    Upon examining additional erroneous cases, we conclude that the feedback component effectively addresses issues such as referencing non-existent functions, using incorrect constants, omitting import statements, and mis-implementing function logic. 

    \subsection{\textbf{RQ3. Accuracy in Refactoring Stage.}}\label{subsection:Eval-RQ3}
    We aim for the refactoring stage to be as accurate as possible to establish a strong foundation for the subsequent generation stage. Thus, we examine the correctness of the refactoring stage within SPDZCoder, and we additionally evaluate its generality across different LLMs. 
    Specifically, we first choose \texttt{GLM-4}, \texttt{DeepSeek-V2.5} and \texttt{DeepSeek-V3} to perform the refactoring task, and then we use \texttt{GPT-4-turbo} to generate five test cases for each Python code sample in SPDZEval, and finally we evaluate the functional correctness of CFPs generated by these models through Python unit tests, with result shown in Table \ref{tab:IRacc}.

    As shown in the table, GPT-4 effectively performs the refactoring task, achieving an overall correctness of 91.24\% in pass@1 and 95.26\% in pass@2. Substituting GPT-4 with weaker models, such as GLM-4 and DeepSeek-V2.5, results in a slight decline in accuracy. For instance, when GLM-4 serves as the backbone model, the overall pass@1 and pass@2 scores drop to 85.77\% and 93.43\%, respectively. Meanwhile, DeepSeek-V3 achieves the same performance as GPT-4, suggesting comparable refactoring capabilities.
\begin{table}[htbp]
  \centering
  \caption{Overall accuracy of refactoring stage using various LLMs as backbone. GPT-4 is empolyed in SPDZCoder.}
  \setlength{\tabcolsep}{15pt}  
  {\renewcommand{\arraystretch}{1.0} 
  \scalebox{0.9}{
    \begin{tabular}{lcc}
    \toprule
    \multicolumn{1}{c}{\multirow{2}[4]{*}{SPDZCoder}} & pass@1 & pass@2 \\
\cmidrule{2-3}          & 91.24 & 95.26 \\
    \midrule
    \textit{with} GLM-4 & 85.77 & 93.43 \\
    \textit{with} DeepSeek-V2.5 & 90.88  & 91.61 \\
    \textit{with} DeepSeek-V3 &  91.24  & 95.26 \\
    \bottomrule
    \end{tabular}%
    }
}
  \label{tab:IRacc}%
\end{table}%



    Along with the results in Table \ref{tab:various-backbone}, we further observe that while using GLM-4 as backbone attains refactoring correctness comparable to GPT-4, its translation performance is not equivalently comparable.  This discrepancy arises because SPDZCoder's translation process requires CFPs to be both functionally correct and structurally compliant with refactoring rules, yet the latter cannot be automatically verified. To investigate further, we manually inspected the CFPs generated by GLM-4 and found that they sometimes did not strictly adhere to the refactoring rule prompts. Common deviations included failing to utilize the provided basic non-linear mathematical functions to rewrite advanced ones and, albeit rarely, violating the obliviousness requirement. This observation empirically suggests that the strong capabilities of in-context learning and instruction following are required for SPDZCoder's backbone.

    \subsection{\textbf{RQ4. Average Token Consumption and the Number of Token Savings by Pattern Match.}}\label{subsection:Eval-RQ4}
    We evaluate the average token consumption of SPDZCoder against baselines, presenting the results in Figure \ref{fig:tokens}.
    As shown in the figure, SPDZCoder does not lead to a significant increase in prompt token consumption compared to API-Doc, the best-performing baseline. Specifically, SPDZCoder consumes $10,343$ prompt tokens and $1,824$ completion tokens per code example. Additionally, removing the repair stage reduces prompt token consumption to $6,377$, which is comparable to API-Doc.

        \begin{figure}[htbp] 
        \centering
        \includegraphics[width=\linewidth]{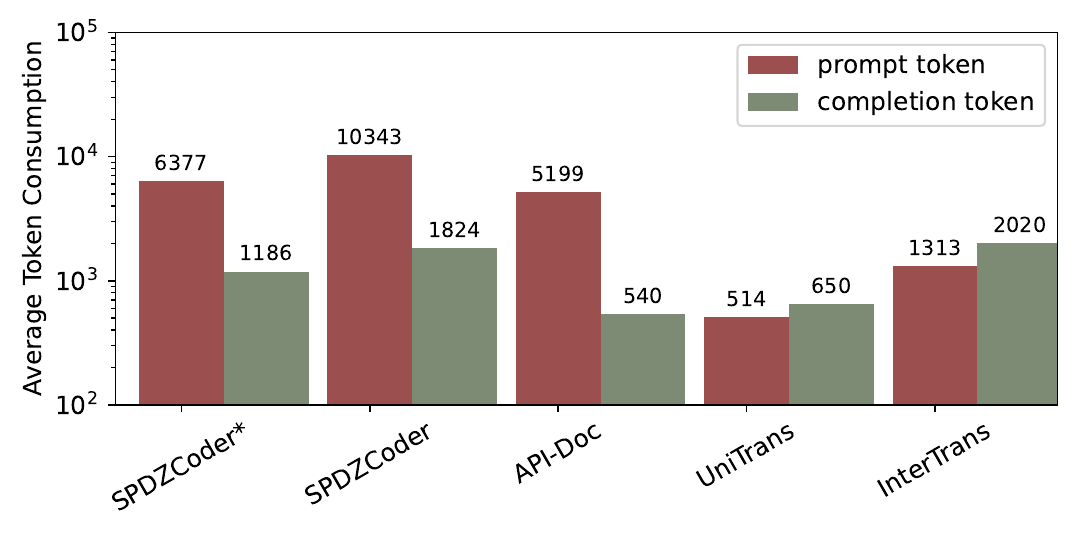} 
        \vspace{-25pt}
          \caption{Average token consumption of SPDZCoder vs. baselines. SPDZCoder* represents the one without repair stage. The y-axis is on a logarithmic scale.}
          \Description[<short description>]{<long description>}
          \label{fig:tokens}
  \end{figure}
    Furthermore, we examine the token savings achieved by our pattern-matching strategies in the refactoring and generation stages. The savings are calculated by subtracting the average token consumption per code example from the total number of tokens in all rule prompts.
    \begin{table}[htbp] 
{\renewcommand{\arraystretch}{1.2} 
\caption{Prompt Token Consumption with/without Pattern Matching Strategies in Refactoring and Generation Stages} 
\centering 
\resizebox{0.5\textwidth}{!}{
    \begin{tabular}{ccc} 
    \hline 
    \bfseries Setting  & \bfseries Refactoring Stage & \bfseries Generation Stage\\ 
    \hline
     \textit{w/o} Pattern Match & 4355 & 8973\\
     \textit{w/}  Pattern Match & 1567 & 4810\\ 
    \hline 
    \end{tabular} 
}
\label{tab:num_saved} 
}
\end{table}
    
    As shown in Table \ref{tab:num_saved}, employing pattern matching significantly reduces prompt token consumption, improving translation efficiency and preventing from applying unnecessary rules. Pattern-matching strategies not only enhance the resource efficiency but also allow us to scale more designed rules within SPDZCoder.

\section{Discussion}

\subsection{Threats to Validity}
\para{Internal Validity.}
The threats to internal validity are as follow: (1) One potential threat is parameter selection, including \texttt{seed}, \texttt{temperature}, and \texttt{top-p}. To mitigate this, we uniformly set these values across all LLMs. While these settings may not be optimal for any specific LLM, they do not impact our ability to analyze performance improvements over direct translation introduced by different methods. 
(2) Data leakage is another possible concern. However, as it affects all evaluated methods equally, the relative comparisons remain valid. Moreover, publicly available training data typically contain limited parallel corpora, particularly when the target programming language involves privacy-preserving computation. 
(3) A third threat is the choice of LLM backbone. While different LLMs may yield varying absolute performance values, the comparative evaluation between SPDZCoder and baseline methods remains unaffected.

\para{External Validity.}
The quality of the evaluation dataset poses a potential threat to external validity. To mitigate this, we selected an undergraduate student with a strong programming background to construct the dataset. The functions in the dataset vary in complexity and diversity, and the final dataset underwent manual inspection by a development engineer. 
Another threat is the choice of experimental models. To address this, we evaluated SPDZCoder across representative general, code-specific, and reasoning LLMs. Additionally, we maintained a consistent prompt template for SPDZCoder across all LLM backbones to ensure fair comparisons.

\subsection{Generalizability of the Approach}
The methodology of SPDZCoder is reusable when the source language changes; however, it requires a new analysis of the semantic-expressing differences between the source language and MP-SPDZ.
Specifically, refactoring rules addressing obliviousness requirements can be directly reused, as these constraints remain consistent across different source languages. In contrast, the remaining refactoring rules and the generation rule depend on the syntactic complexity of the source language, making them less transferable without modification.

\section{Related Work}
\subsection{Large Language Models}
General LLMs \cite{brownLanguageModelsAre2020a, chowdhery_palm_2022, touvron_llama_2023, touvron_llama_2023-1, claude3, openai_gpt-4_2023, gemini_team_gemini_2023, workshop_bloom_2023, jiang_mixtral_2024, glm2024chatglm, gpt-4o, deepseekai2024deepseekv3technicalreport} have rapidly advanced, demonstrating remarkable capabilities across diverse tasks. For example, GPT-4o has achieved state-of-the-art performance on code benchmarks\cite{chen_evaluating_2021, austin_program_2021}. 
Building on the success of LLMs, many research efforts have leveraged their capabilities to address specific challenges in software engineering \cite{yang_large_2024, li_fault_2022, ma_knowlog_2024, deng_large_2024, huang_crashtranslator_2024}, including code translation \cite{yangExploringUnleashingPower2024, macedoInterTransLeveragingTransitive2024a} (discussed below).

Code LLMs are trained on multiple programming languages to support code understanding and generation.
A collection of studies focuses pre-training code LLMs, such as  DeepSeek-V2.5~\cite{deepseekv2}, Qwen2.5-Coder~\cite{hui2024qwen2}, CodeGen~\cite{nijkamp_codegen_2023}, CodeGeeX~\cite{zheng_codegeex_2023}, StarCoder~\cite{li_starcoder_2023}, and DeepSeek-Coder~\cite{guo_deepseek-coder_2024}. 
Another line of research focuses on fine-tuning them to improve performance or to adapt them to specific coding tasks, such as WizardCoder\cite{luo_wizardcoder_2023} and InstructCodeT5+~\cite{wang_codet5_2023}.

Reasoning LLMs enhance LLMs' capability in structured logical inference, which is particularly crucial for coding and mathematical tasks.
DeepSeek-R1 \cite{deepseekai2025deepseekr1incentivizingreasoningcapability} is trained via large-scale reinforcement learning (RL) as the preliminary step, and it achieves performance comparable to OpenAI-o1 on reasoning tasks. Qwen-QwQ \cite{qwq-32b-preview} and OpenAI o1 \cite{openai-o1} also belong to RLLMs.

Pre-training or fine-tuning LLMs requires extensive code, which is scarce in privacy computing. Instead, we employ in-context learning to enhance code translation.

\subsection{Code Translation}

Code translation approaches generally fall into three categories: transpiler-based, learning-based and LLM-based methods. 
Transpiler approaches \cite{noauthor_java_nodate, noauthor_c2rust_nodate, noauthor_cxgo_nodate} rely on program analysis techniques to convert code between languages but are limited to syntactic transformations.
Learning-based methods leverage Transformer-based architectures, requiring training data to develop deep models for either specific or general code translation tasks. Notable works in this category include CodeBERT\cite{feng_codebert_2020}, CuBERT\cite{kanade_learning_2020}, GraphCodeBERT\cite{guo_graphcodebert_2021}, CodeT5\cite{wang_codet5_2021} and PLBART\cite{ahmad_unified_2021}.

Beyond deep learning, LLMs are employed as backbone to perform code translation. For example, UniTrans \cite{yangExploringUnleashingPower2024} incorporates test cases to augment code translation and iteratively repairs bugs for incorrectly translated programs prompted by test case execution results. 
InterTrans \cite{macedoInterTransLeveragingTransitive2024a} utilizes a planning algorithm (ToCT) to generate candidate translation paths, and leverages intermediate translations to enhance code translation. SPDZCoder also uses LLMs for code translation, but unlike UniTrans and InterTrans, our target programming language is not familiar to LLMs.

\subsection{Code Generation Benchmark}
Many works were proposed to benchmark the code generation ability of LLMs \cite{chen_evaluating_2021,liu_is_2023,yu_codereval_2024,athiwaratkunMultilingualEvaluationCode2022,lai2022ds1000naturalreliablebenchmark,cassanoMultiPLEScalablePolyglot2023,hendrycksapps2021,austin_program_2021,du_classeval_2023}. The Mostly Basic Programming Problems (MBPP) dataset \cite{austin_program_2021}, consisting of 974 programming tasks, measures the ability of LLMs to synthesize short Python programs from natural language descriptions. 
Human-eval\cite{chen_evaluating_2021} measures functional correctness for synthesizing programs from docstring.
Liu et al. proposed an enhanced code synthesis evaluation framework named EvalPlus\cite{liu_is_2023}, which augments HumanEval with large amounts of newly produced test-cases. Similar work also includes Multi-HumanEval \cite{athiwaratkunMultilingualEvaluationCode2022}.
More recently, parallel with ClassEval\cite{du_classeval_2023}, Yu et al. proposed CoderEval\cite{yu_codereval_2024} to evaluate the effectiveness of code LLMs in generating code beyond only standalone functions.
 There are also code generation benchmarks for specific purpose, \eg type inference \cite{mir_manytypes4py_2021}. 
SPDZEval differs from them in terms of coding task and target programming language.

\section{Conclusion}
We propose SPDZCoder, a rule-based framework designed to teach LLMs to harness code translation in privacy computing settings. 
SPDZCoder utilizes expert-defined rules to help LLMs learn the unique ways and constraints of expressing common semantics in MP-SPDZ, without requiring experts to manually generate massive amounts of training data. 
It addresses this task by considering semantic-expressing differences at various levels, leveraging the refactoring and generation stages to mitigate these differences at high and low levels, respectively. Furthermore, SPDZCoder incorporates execution messages from test cases as feedback, guiding LLMs to rectify incorrectly translated programs. 
To evaluate SPDZCoder against recent advanced baselines, we introduce SPDZEval, a benchmark dataset tailored for this task. 
Extensive experimental results demonstrate that SPDZCoder significantly outperforms strong baselines and maintains consistent effectiveness across different LLM backbones.
Our work provides new insights into teaching LLMs to synthesize privacy computing code and suggests that SPDZCoder can facilitate better code translation practices within the privacy computing community. 
For future work, we aim to synthesize privacy computing code directly from natural language.



\bibliographystyle{ACM-Reference-Format}
\bibliography{ICSE_biblatex-rebiber0731}

\clearpage 
\appendix
\section{Supplementary Material}
\subsection{Translation Example}
\label{app:translation-example}
We provide qualitative examples of translation in Figure \ref{fig:translation-example} to better understand how the rules works.

\subsection{Prompt Template}
\label{app:prompt-template}
SPDZCoder utilizes a set of prompting templates to guide LLMs in generating MP-SPDZ code. Below, we present representative prompt templates: (1) an example of high-level rule  \texttt{EliminateBreak}, which instructs LLM to eliminate break statement (Table \ref{tab:table1}); (2) the low-level rule, which prompts LLMs to generate MP-SPDZ code (Table \ref{tab:table2}); (3) the self-reflection rule (Table \ref{tab:table3}) to alleviate hallucination; (4) the repair rule to prompt LLMs to rectify bugs or incorrect logic in generated MP-SPDZ code (Table \ref{tab:table4}).

\subsection{Prompt Template for API Doc Baseline}
\label{app:prompt-baseline}
We show the complete prompt template for API Doc method in Table \ref{tab:table5}.

    \begin{figure*}[htbp] 
        \centering
        \includegraphics[width=\textwidth]{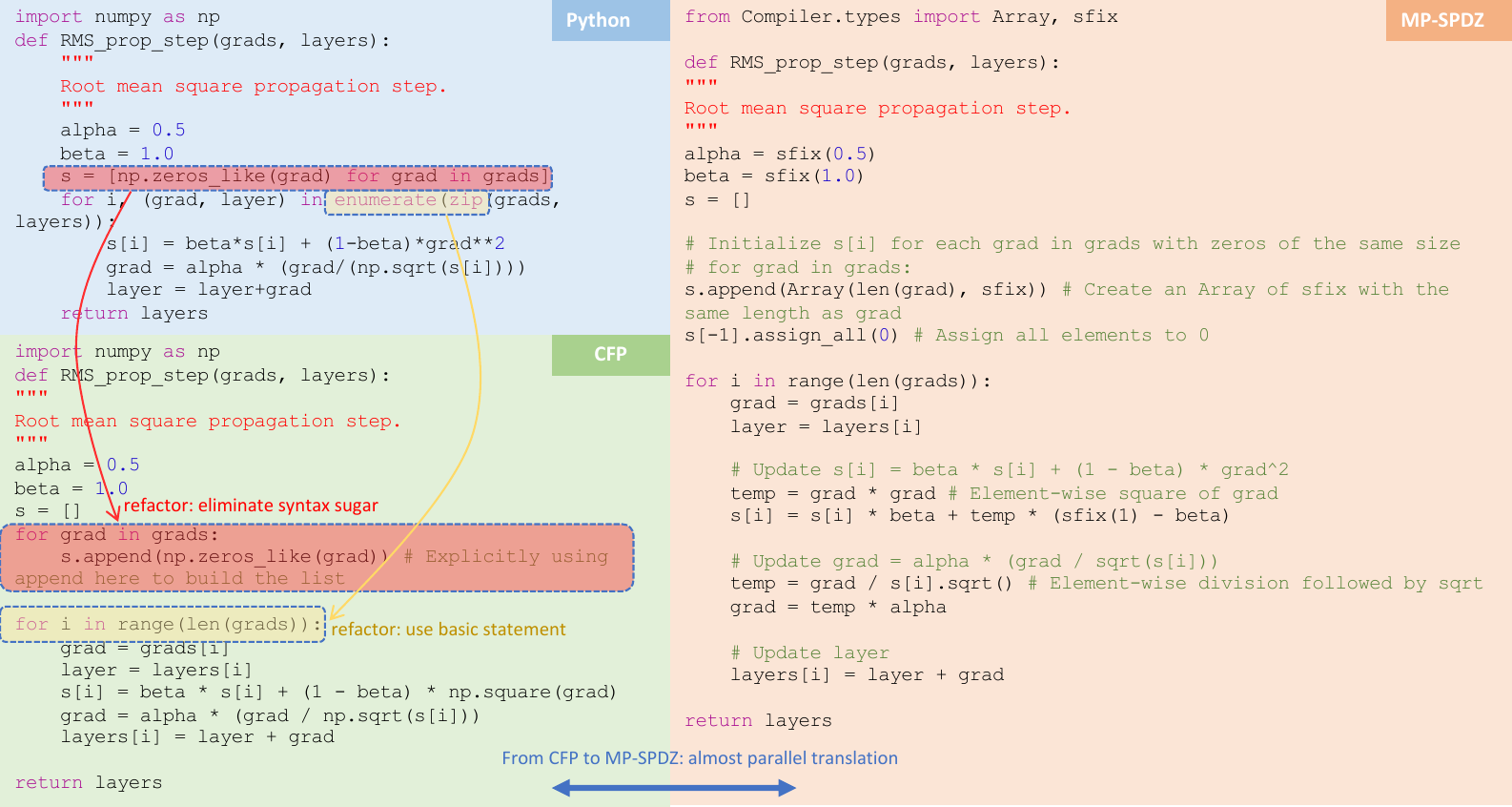} 
          \caption{A translation example of SPDZCoder}
          \Description[<short description>]{<long description>}
          \label{fig:translation-example}
  \end{figure*}

\begin{table*}[htbp]
    \centering
    \begin{tabular}{|p{0.98\textwidth}|}
          \hline
          \vspace*{0.1cm}
          \textbf{System Prompt:} \\
          \smallskip
          You are an expert to write python but you always write python through a basic and explicit approach and do not use `break` and `continue` keyword in order that junior student can understand it. \\
          \\
          \textbf{User Prompt:}
        \begin{verbatim}
Refactor the following code in order that the `break` keyword is eliminated. 
You are not allowed to use any `break` or `continue` keyword.
```
{CODE}
```
You can refer the given example:
```
for i in range(len(array)):
    if a[i]>10:
        if a[i+1]>2:
            break       # The condition for `break` is `a[i]>10 and a[i+1]>2`
        else:
            a[i] += 1   # The condition for `a[i] += 1` is `a[i]>10 and a[i+1]<=2`
    else
        a[i] += 2       # The condition for `a[i] += 2` is `a[i]<=10`
```
will be refactored into
```
# Employ a flag to manage the loop execution in order to 
# avoid using both `break` and `continue` keyword
flag = False

for i in range(len(array)):
    flag = flag or (a[i]>10 and a[i+1]>2)
    # The result is an combination
    a[i] = flag*a[i] + (1-flag)*((a[i]>10 and a[i+1]<=2)*(a[i]+1) + (a[i]<=10)*(a[i]+2))
    # The code is in an oblivious form since there is no `break` 
    # and `continue` (no jump) statement in the code.
```
        \end{verbatim} \\
    \vspace{-10pt}
    \textbf{Dataclass:} 
    \begin{verbatim}
@dataclass
class EliminateBreak:
    system_prompt: str
    rule_prompt: str
eliminate_break = EliminateBreak(System_prompt, User_prompt) 
    \end{verbatim} \\
    \hline
    \end{tabular}
    \caption{Prompt template for \texttt{EliminateBreak} rule, which will eliminate \texttt{break} statement in Python code.} 
    \label{tab:table1}
\end{table*}


\newpage
\begin{table*}[thbp]
    \centering
    \small
    \begin{tabular}{|p{0.98\textwidth}|}
          \hline
          \vspace*{0.1cm}
      \textbf{System Prompt:} \\
          \smallskip
          You are an expert to write MP-SPDZ promgram and you are familar with the differences between Python and MP-SPDZ. When you translate a Python Program into MP-SPDZ, you will carefully consider their differences in expressing semantics and try to solve the code translation task line by line and step by step. \\
          \\
      \textbf{User Prompt 1:} \\
        Can you translate Python code into MP-SPDZ code? \\
      \textbf{Assistant Response 1:} \\
        Yes, I can help you translate Python code into MP-SPDZ code, but the process depends on \dots 
        Could you provide the Python code or describe the functionality you are interested in? \\
        \smallskip
      \textbf{User Prompt 2:} \\
        What do you think you should take care when you perform such translation task? In other words, how do you finish translating Python to MP-SPDZ step by step? Give your answer in detail. \\
      \textbf{Assistant Response 2:} \\
        Translating Python code into MP-SPDZ code requires careful consideration of several factors unique to the secure multi-party computation (MPC) environment of MP-SPDZ. Here's a step-by-step guide to ensure a proper translation:
        
        Step 1: Understand the Functional Requirements.
        First, \dots 
        
        Step 2: Analyze Data Sensitivity and Security Requirements. Determine \dots 
        
        Step 3: Map Python Functions to MP-SPDZ Functions. Identify equivalents or approximations for Python functions and libraries in MP-SPDZ: \\
        - Basic operations like addition, subtraction, multiplication, and division usually have direct equivalents.\\
        - More complex functions, such as 
        \dots might involve using specific MPC protocols supported by MP-SPDZ.\\
        \dots \\
        Step 5: \dots  
        
        
      \dots \\
      \textbf{User Prompt 5:} 
        \begin{verbatim}
Given a Python code snippet, your task is to understand the semantic of the Python code and translate 
it into MP-SPDZ code....The following are some typical  math functions in MP-SPDZ `mpc_math` module 
(`from Compiler import mpc_math`) for your reference, always keep them in your mind!!!
For nonlinear arithmetic functions:
`mpc_math.exp2_fx(x)` calculates `2^x` i.e. `numpy.exp2(x)`;
`mpc_math.log2_fx(x)` calculates base-2 logarithm of x `numpy.log2(x)`;
...
For trigonometric functions, `sin`, `cos`, `tan`... in Python are `mpc_math.sin`, `mpc_math.cos`, 
`mpc_math.tan ... in MP-SPDZ respectivly. Also, you should import necessary modules first, 
otherwise I will punish you:
```
import math
from Compiler import mpc_math
```
Finally, remember that logarithms have several key properties that make them incredibly useful ....
Now let us start to translate the given python code below:
```python
def logaddexp2(x1, x2):
    import numpy
    # Convert `logaddexp2` to fundamental operations using `exp` and `ln`
    # First, convert inputs from base-2 log to natural log(e) by multiplying with ln(2)
    ln2 = numpy.log(2)      # compute natural log of 2 i.e. log(2)
    x1_ln = x1 * ln2
    x2_ln = x2 * ln2
    # Calculate the exponentials of the inputs
    ......
    result = numpy.log(sum_exp) / ln2   # first compute natural log of sum_exp and then divide by ln(2)
    return result
```
        \end{verbatim}
        \vspace{-10pt}
      \textbf{Assistant Response 5:} \dots \\
      \\
      \textbf{User Prompt :}
        \begin{verbatim}
Your task is to understand each line of code in the given Python/Numpy program and translate it into 
semantically equivalent MP-SPDZ code..... Thus, it is very important to refer to the chat history to 
get the correct/useful usage of MP-SPDZ methods/functions and never add any usage example. 
Now, carefully translate the following Python code in MP-SPDZ code.
{CODE}
        \end{verbatim} \\
    \hline
    \end{tabular}
    \vspace*{-10pt}
    \caption{The in-context learning demonstrations for code generation, which are selected by pattern match and organized into an LLM chat message.}
    \label{tab:table2}
\end{table*}


\newpage
\begin{table*}[htbp]
    \centering
    \begin{tabular}{|p{0.98\textwidth}|}
          \hline
          \vspace*{0.1cm}
          \textbf{System Prompt:} \\
          \smallskip
          Correct/rectify the given MP-SPDZ program with given instruction if applicable. \\
          \\
          \textbf{User Prompt:}
        \begin{verbatim}
Review the following MP-SPDZ program and follow the given instructions as below:
1. Rectify those incorrectly imported modules. Here are the correct examples to import 
Python and MP-SPDZ modules. If `mpc_math` is used, never forget to import it!
```
# import math related module
import math
from Compiler import mpc_math

# import type related module
from Compiler.types import sint
from Compiler.types import sfix
from Compiler.types import cint
from Compiler.types import cfix
from Compiler.types import regint
from Compiler.types import Array
from Compiler.types import Matrix
from Compiler.types import MemValue

# import all modules from standard library (optional)
from Compiler.library import *
```

2. Rectify non-exist MP-SPDZ Functions.
- `mpc_math.exp(x)` should be `mpc_math.pow_fx(math.e, x)` which computes `e^x`
- `mpc_math.log(x)` should be `mpc_math.log_fx(x, math.e)` which computes `ln(x)`
- `mpc_math.log_fx(x, cfix(math.e)) should be `mpc_math.log_fx(x, math.e)` 
   which computes `ln(x)`
- `mpc_math.sqrt_fx(x)` should be `mpc_math.sqrt(x)` which computes `sqrt(x). 
   Before computing square root, convert `x` into `sfix` data type by `x = sfix(x)`.
- `mpc_math.pi` should be `sfix(mpc_math.pi)`, providing a fixed-point approximation of pi.
- `math.pi` should be `sfix(math.pi)` which provides a fixed-point approximation of pi.
- `mpc_math.pi_fx()` should be `sfix(mpc_math.pi)` or `sfix(math.pi)` 
   which provides a fixed-point approximation of pi.

3. Delete/Remove the part of `example usage of the function` in the code if applicable.
4. Tenary expression `x if condition else y` should be `condition.if_else(x,y)`.
5. `mpc_math.max(y, 0)` should be `y.get_vector().max(0)`.

Strictly follow the above 5 aspects and start to review the code. If applicable, 
return the modified code, otherwise return the original code as your response.
```MP-SPDZ
{CODE}
```
        \end{verbatim} \\
    \hline
    \end{tabular}
    \caption{Prompt template for Self-reflection}
    \label{tab:table3}
\end{table*}


\newpage
\begin{table*}[htbp]
    \centering
    \begin{tabular}{|p{0.98\textwidth}|}
          \hline
          \vspace*{0.1cm}
          \textbf{System Prompt:} \\
          \smallskip
          You are an expert to debug/re-write MP-SPDZ code when you are given the traceback information/compilation errors/ runtime errors. \\
          \\
          \textbf{User Prompt:}
        \begin{verbatim}
You are provided a Python code snippet as follows:
```python
{PYTHON_CODE}
```
Then we provide its corresponding semantic-equivalent MP-SPDZ code in the following:
```MP-SPDZ
{SPDZ_CODE}
```
However, there exists bugs/errors in the provided MP-SPDZ code 
and the traceback inforamtion is provided to you in the below:
"
{COMPILATION_RUNTIME_ERROR}
"
Here is your task: First, read the Python code and the MP-SPDZ code. 
Then, combining traceback inforamtion, you fix the existing bugs in the given MP-SPDZ code 
or re-translate the Python code into MP-SPDZ again. 
Finally, return your rectified/correct MP-SPDZ code. 
        \end{verbatim} \\
    \vspace{-10pt}
***********************************************************************************************\\
    \vspace{1pt}
    \textbf{System Prompt:} \\
    \smallskip
    You are an expert to translate Python code into MP-SPDZ code and refine MP-SPDZ code. \\
    \\
    \textbf{User Prompt:} \\
    \vspace{-10pt}
        \begin{verbatim}
You are provided a Python code snippet as follows:
```python
{PYTHON_CODE}
```
Then we provide you a MP-SPDZ code snippet in the following and we hope the given MP-SPDZ
code performs the same functionality as the Python code, i.e. semantically equivalent
to the given Python Code:
```MP-SPDZ
{SPDZ_CODE}
```
However, the provided MP-SPDZ code performs wrong functionality. The reason is that
the code translation task from Python to MP-SPDZ failed.
To this end, your task is: 
First, read the Python code, and learn/summarize its functionality. Then, 
you fix the given MP-SPDZ code or totally re-translate the Python code into MP-SPDZ again. 
Finally, return your rectified MP-SPDZ code with correct functionality. 
        \end{verbatim} \\
    \hline
    \end{tabular}
    \caption{Prompt template for Repair Stage}
    \label{tab:table4}
\end{table*}
\clearpage
\begin{table*}[htbp]
    \centering
    \begin{tabular}{|p{0.98\textwidth}|}
          \hline
          \vspace*{0.1cm}
          \textbf{System Prompt for Python Code Summarization:} \\
          \smallskip
          You are an expert to write Python, and you are good at describing/explaining a Python program to other people. \\
          \\
          \textbf{User Prompt for Python Code Summarization:}
        \begin{verbatim}
Following is a piece of python code and some annotations about it. Your task is to describe 
the semantic of the code, i.e., you describe/explain the functionality of the code in 
natural languge within 5 sentences. For functions defined in the code, you should summarize
its inputs, outputs and functionalities. For cirtical global variables in the code, you 
should summarize their names and usages.

Let us try with the following python code snippet.
```
{CODE}
```
        \end{verbatim} \\
    \vspace{-10pt}
***********************************************************************************************\\
    \vspace{1pt}
    \textbf{System Prompt for MP-SPDZ Code Generation:} \\
    \smallskip
    You are an expert to write MP-SPDZ program and you are familar with the differences between Python and MP-SPDZ. Thus, you think carefully before you write MP-SPDZ program. In addtion, you are very smart to refer the above/aforementioned external knowledge. \\
    \\
    \textbf{User Prompt for MP-SPDZ Code Generation:} \\
    \vspace{-10pt}
        \begin{verbatim}
This the MP-SPDZ API document:
{API_DOC}

The following is a python function and its semantic description. Your task is to 
implement/write a piece of code under MP-SPDZ framework according to the description. 
You should refer to the given MP-SPDZ API document above and write the code carefully.
The Python code is as follows:
```python
{CODE}
```
The semantic description of the code is here:
"
{DESCPRIPTION}
"
When you implement/write the MP-SPDZ code, you are supposed to follow the below requirements:
1. The code you write must have the same functionality as the original code. 
   The critical parameters or variables must keep the same name.
2. You should use the types and methods of the MP-SPDZ framework correctly to rewrite the code.
   For example, you should change the `list` type into `Array` type.
3. All variables should be viewed as ciphertext variables, and you should turn them into 
   secret types in MP-SPDZ and should not reveal them.
4. You only need to guarantee the functionality of the code you write matches the input code, 
   and you don't have to align the implementation between the input and your answer.
        \end{verbatim} \\
    \hline
    \end{tabular}
    \caption{Prompt templates for API Doc Baseline}
    \label{tab:table5}
\end{table*}

\end{document}